\documentclass{emulateapj}
\usepackage{natbib,graphicx}

\slugcomment{To appear in ApJ}

\shorttitle{Matryoshka Holes in the Transitional Disk Oph IRS 48}
\shortauthors{Brown et al.}

\begin{document}


\title{Matryoshka Holes: Nested Emission Rings in the Transitional Disk Oph IRS 48}


\author{{J.M. Brown\altaffilmark{1}, K.A. Rosenfeld\altaffilmark{1}, S.M. Andrews\altaffilmark{1}, D.J. Wilner\altaffilmark{1}, E.F. van Dishoeck\altaffilmark{2,3}
}}
\altaffiltext{1}{Harvard-Smithsonian Center for Astrophysics, 60 Garden St., MS 78, Cambridge, MA 02138; joannabrown@cfa.harvard.edu}
\altaffiltext{2}{Leiden Observatory, P.O. Box 9513, NL-2300 RA Leiden, The Netherlands}
\altaffiltext{3}{Max-Planck-Institut f{\"u}r extraterrestrische Physik, Postfach 1312,  85741 Garching, Germany}



\begin{abstract} 
The processes that form transition disks - disks with depleted inner
regions - are not well understood; possible scenarios include planet
formation, grain growth and photoevaporation. Disks with spatially
resolved dust holes are rare, but, in general, even less is known
about the gas structure. The disk surrounding the A0 star Oph IRS 48
in the nearby $\rho$ Ophiuchus region has a 30 AU radius hole
previously detected in the 18.7 $\mu$m dust continuum and in warm CO
in the 5 $\mu$m fundamental ro-vibrational band. We present here
Submillimeter Array 880 $\mu$m continuum imaging resolving an inner
hole. However, the radius of the hole in the millimeter dust is only
13 AU, significantly smaller than measured at other wavelengths. The
nesting structure of the disk is counter-intuitive, with increasingly
large radii rings of emission seen in the millimeter dust (12.9 $^{\rm
+1.7}_{\rm -3.4}$ AU), 5 $\mu$m CO (30 AU) and 18.7 $\mu$m dust
(peaking at 55 AU). We discuss possible explanations for this
structure, including self-shadowing that cools the disk surface
layers, photodissociation of CO, and photoevaporation. However,
understanding this unusual disk within the stringent multi-wavelength
spatial constraints will require further observations to search for
cold atomic and molecular gas.

\end{abstract}




\section{Introduction}

Understanding the interplay between dust and gas is crucial for
determining the evolutionary pathways of protoplanetary disks. Gas
dominates the mass and dynamics of the disk, but dust is both easier
to trace observationally and contributes directly to planet core
formation. Models predict that massive planets clear their orbital
paths, creating distinctive gaps in their natal disks \citep{crida07,
bryden99, artymowicz94}. Transition disks, characterized by observed
dust deficits in the inner regions, may be the result of this process
(e.g. \citealt{strom89, calvet02, brown07, najita07}). The recent
discoveries of young substellar candidates in the transition disks T
Cha and LkCa 15 lend credence to this hypothesis
(\citealt{huelamo11,kraus12}). However, processes other than orbital
clearing can potentially cause dust deficits, including
photoevaporation and grain growth. The gas distribution could be
diagnostic of different scenarios. In disk photoevaporation models,
the dust is tied to the gas resulting in similar distributions and
timescales \citep{alexander06,owen11}. In contrast, grain growth to
very large sizes would result in an apparent dust hole but no drop in
the gas density. Finally, stellar and planetary companions present an
intermediate case with some movement of gas across the gaps into the
inner disk expected depending on companion mass and location
\citep{artymowicz96,zhu11}. The presence of gaps can enhance other
disk clearing mechanisms such as the magneto-rotational instability
\citep{chiang07} and photoevaporation \citep{alexander09}. Knowledge
of the gas distribution is also particularly relevant for studying
young planets as the gas drives both formation and migration.

The A0 star IRS 48 in the $\rho$ Ophiuchi star formation region
($\sim$ 1 Myr, \citealt{luhman99}; 121 pc, \citealt{loinard08}) has an
unusual disk. Spatially resolved 18.7 $\mu$m imaging of the dust
continuum revealed a ring-like structure with a peak-to-peak diameter
of 110 AU and a central hole with a radius of $\sim$30 AU
\citep{geers07}. Strong polycyclic aromatic hydrocarbon (PAH)
emission, commonly a tracer of gaseous molecular material, was
unresolved and centered on the stellar position, interior to the ring
seen in the 18.7 $\mu$m dust. The CO gas might be expected to follow
the PAH distribution and thus be located inside the dust
hole. However, resolved imaging with the VLT-CRIRES showed warm CO
emission in a 30 AU radius ring, coincident with the proposed dust
hole wall \citep{brown12}. The presence of PAHs was explained by their
longer survival time relative to CO in high UV flux, low density
regions. \citet{brown12} suggest that truncation of the gas and dust,
as might be caused by a planet, lowers the density in the inner region
leading to photodissociation of CO interior to the truncation
radius. This scenario should lead to a dust hole detectable with
millimeter interferometry.

Millimeter imaging of transitional disks provides some of the best
constraints on cavity size and characteristics. Resolving the cavities
requires both long baselines and good (u, v) coverage to achieve the
necessary high spatial resolution and dynamic range. The optically
thin emission observed in the submillimeter is very sensitive to the
mass surface density profile. However, spatially resolved studies have
been restricted to the brightest disks. Despite improving statistics,
only a limited number of transition disks have been observed at the
spatial scales necessary to resolve the inner cavity
(\citealt{pietu06}, \citealt{hughes07, hughes09}, \citealt{brown08,
brown09}, \citealt{andrews09, andrews11}).

We present in this paper Submillimeter Array (SMA) 880 $\mu$m imaging
showing a hole in the distribution of large dust grains in the disk
around Oph IRS 48. However, surprisingly, the size of the millimeter
dust hole is significantly smaller than that observed in the 18.7 $\mu$m
dust and 5 $\mu$m gas. We discuss the observations (\S \ref{obs}), the
models used to determine the hole size (\S \ref{analysis}) and finally
discuss the implications of the millimeter data and some theories as
to the cause of the discrepancy (\S \ref{discuss}).

\section{Observations and Results}
\label{obs}

We observed Oph IRS 48 (WLY 2-48) with the Submillimeter Array (SMA)
  in the very extended configuration at 345 GHz (880 $\mu$m). Two
  tracks were obtained with eight antennas on 2011
  August 29 and five antennas on 2011 August 26. The position
  measured by the SMA was $\alpha$ = 16$^{\rm h}$27$^{\rm m}$37$^{\rm
  s}$.17, $\delta$ = -24$^{\rm \circ}$30$^{\rm '}$35$^{\rm ''}$.5,
  J2000. Baseline lengths ranged from 30 to 590 m.  The beam size was
  0\farcs45 by 0\farcs24 arcsec at a position angle of
  27.4$^\circ$. Double sideband (DSB) receivers tuned to 341.165 GHz
  provided 4 GHz of bandwidth per sideband.  The visibility phases
  were calibrated using the quasars J1625-254 (1.4 Jy)
  and J1626-298 (0.8 Jy), with 10 minute cycles. 3C454.3 was used for
  passband calibration. The absolute flux scale was calibrated from a
  measurement of Neptune from the same night. The uncertainty in the flux scale
  is estimated to be 15\%.

Figure \ref{fig:uv} presents the image (left) and visibilities
(right).  While the image is clearly not centrally peaked, we focus
the analysis on the visibilities where a null more clearly defines the
structure change. Sharp transitions in the mass surface density, such
as a hole, are seen in aperture synthesis observations as a null in
the flux distribution with (u,v) distance (e.g.,
\citealt{andrews11}). In contrast, most classical T Tauri stars have
power-law mass surface density profiles, resulting in a smooth decline
of flux with (u,v) distance. Another advantage of visibility domain
analysis is a better understanding of the uncertainties, due to the
lack of additional image processing, including Fourier transformation
and nonlinear deconvolution.  The visibilities in Figure 1 were
deprojected to account for position angle, PA, and disk inclination,
i, and binned assuming a circular disk (e.g., \citealt{hughes07}). The
deprojected (u,v) distance is $R = \sqrt{d_{\rm a}^2 + d_{\rm b}^2}$
with $d_{\rm a}= R {\rm \, sin} \phi$ and $d_{\rm b}= R{\rm \, cos}
\phi \, {\rm cos}\, i$ where $\phi={\rm arctan(v/u)}$ - PA.

Earlier archival data from 2007 May 14 observed IRS 48 at 230 GHz (1.3
mm) in the very extended configuration. The track had only 6 working
antennas and a beam size of 0\farcs67x0\farcs4 (P.A. =
-7.8$^\circ$). The spatial resolution was inadequate to clearly
resolve the millimeter dust hole, but the data indicated that the
millimeter dust was unlikely to follow the shorter wavelength
emission. The visibilities are overlaid in gray on Figure \ref{fig:uv}
but all further analysis uses the higher spatial resolution 880 $\mu$m
data.

Spatial filtering of flux from the outer disk beyond $\sim$100 AU is a
concern with the long baselines in the SMA data. IRS 48 has been
observed with single dish millimeter telescopes with a flux of
180$\pm$9 mJy with JCMT-SCUBA at 860 $\mu$m \citep{andrews07} and
60$\pm$10 mJy with the IRAM 30m at 1.3 mm \citep{motte98}. The SMA
tracks have total fluxes of 160 mJy at 880 $\mu$m and 50 mJy at 1.3
mm. Less than 20\% of the total flux has thus been lost in the
interferometric data, ruling out the possibility of a large reservoir
of outer disk millimeter dust.

\section{Modeling}
\label{analysis}

\begin{figure*}[t!]
\begin{center}
\includegraphics[angle=0,scale=.9]{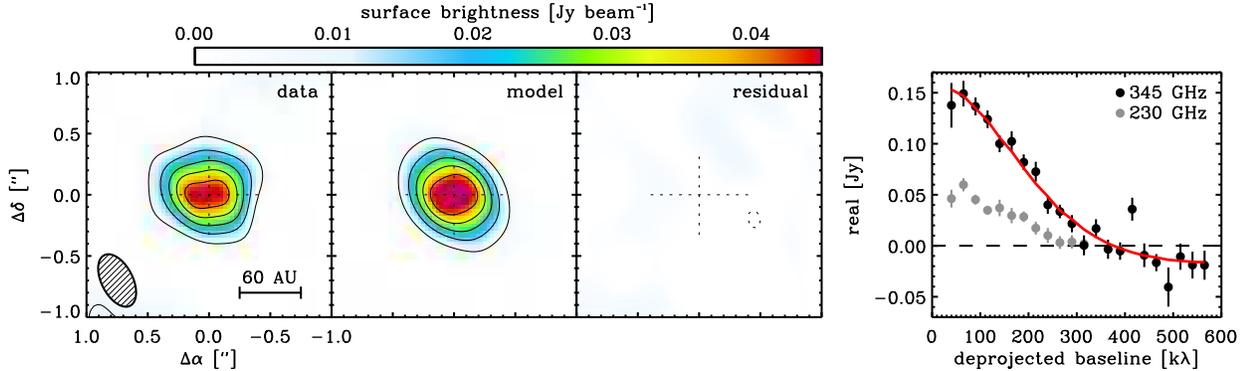} 
\end{center}
\caption{Images of the data (left), model (center left) and residuals
(center right). The contours are 3$\sigma$ intervals where the RMS
noise is 2.7 mJy/beam. The (u,v) visibities for 345 GHz (black) and
230 GHz (gray) are on the far right with the best fit model from Table
\ref{table:model} in red.  \label{fig:uv}}
\end{figure*}

We model the millimeter emission following the similarity solution
approach which has been used to characterize the millimeter dust in
both normal and transition disks \citep{hughes08, andrews09,
andrews11b, andrews11}. Our emission model is taken to be azimuthally
symmetric with a surface brightness profile appropriate for optically
thin thermal emission,
\begin{equation}
I_\nu \propto B_\nu(T_d)(1 - e^{-\tau}) \approx B_\nu(T_d) \tau 
\end{equation}
where $B_\nu$ is
the Planck function, $T_d$ is the dust temperature, and $\tau$ is the
optical depth. The dust temperature is assumed to follow a power-law
profile normalized by the stellar temperature of IRS 48: 
\begin{equation}
T_d = T_* (\frac{r}{r_*})^{-0.5}
\end{equation}
where $T_*$=9000 K and $r_*$=1.35 R$_\odot$. The optical depth across
the disk is set by the surface density for which we adopt the
similarity solution appropriate for a viscous accretion disk
\citep{lynden-bell74, hartmann98}:
\begin{equation}
\tau \propto \left(\frac{r}{r_c}\right)^{-\gamma } \exp {\left[-\left(\frac{r}{r_c}\right)^{2-\gamma }\right]}
\end{equation}
where $\gamma$ sets the gradient of the profile and $r_c$ is the
characteristic scaling radius. We include a cavity by setting to zero
any emission within its radius, $R_{\rm cav}$.

After normalizing and projecting the model image onto the sky plane
with some inclination, position angle, and positional offsets, we use
MIRIAD to sample the model visibilities. The position angle is fixed
to 90$^\circ$ based on the shorter wavelength data. The fitness of
each model, or the likelihood, is constructed from the chi-squared
statistic assuming the visibilities have uncorrelated, Gaussian
errors, $\mathcal{L} \propto \exp (-\chi^2/2)$. To estimate the best
fit model parameters and their associated uncertainties, we sample the
posterior distribution using an ensemble MCMC algorithm
\citep{goodman10, foreman-mackey12}. We report the best fit parameters
and the uncertainties from the 68\% confidence intervals of the
marginalized distribution (Table \ref{table:model}). The best fit
model image and visibility fits are shown in Fig. \ref{fig:uv}.

We also explored a couple of alternate models to understand the
constraints the data put on the millimeter dust distribution. First, a
constant density ring model required flux between 7 and 48 AU to match
the visibilities. Second, $R_{\rm cav}$ was fixed at 30 AU to match
the warm CO gas ring \citep{brown12} and a constant density of dust
was introduced interior to $R_{\rm cav}$. This model still required
50\% of the millimeter flux to come from within $R_{\rm cav}$. In both
cases, the fit produced looked adequate by eye but had a larger
$\chi^2$ than the similarity solution fit reported in Table
\ref{table:model}. The implication is that, regardless of model
details, both a central dropoff in the millimeter emission and substantial dust
within the 5 $\mu$m CO ring are required.

Both the 18.7 $\mu$m and CO images show asymmetries with the north
side brighter than the south. However, the millimeter image shows no
significant asymmetries beyond those expected from an inclined
disk. The asymmetries at shorter wavelengths may reflect variable
extinction of the illuminating radiation field.

The derived inclination of 35$^\circ$ is smaller than those calculated
from the 5 $\mu$m CO (42$^\circ \pm$6$^\circ$) and the 18.7 $\mu$m
dust (48$^\circ \pm$8$^\circ$). The inclination was left as a free
parameter in the similarity solution models with lower inclinations
resulting in a smoother visibility profile and smaller $\chi^2$. This
could be the result of azimuthal averaging over asymmetric structure
that is unseen at the SMA resolution. Large inclinations result in a
slightly larger central cavity in the millimeter dust with 45$^\circ$
corresponding to a best fit cavity size of 15 AU.

\begin{deluxetable}{lccc}
\tablecolumns{4}
\tablewidth{0pt} 
\tabletypesize{\normalsize}
\tablecaption{\label{table:model}Model parameters}
\tablehead{\colhead{Parameter} & \colhead{Best fist} & \colhead{Lower bound} & \colhead{Upper bound}} 
\startdata
$F_{\rm 880 \mu m}$ (Jy) &     0.16 &   0.150  &   0.165   \\
$r_c$ (AU)         &       35 &      27  &      36  \\
$\gamma$           &     0.12 &   -0.49  &     0.6  \\
$R_{\rm cav}$ (AU) &     12.9 &     9.5  &    14.6  \\
Inclination ($^\circ$) &   35 &      28  &      38 
\enddata
\end{deluxetable}

\begin{figure}[t!]
\hspace{-1cm}
\begin{center}
\includegraphics[angle=90,scale=.35]{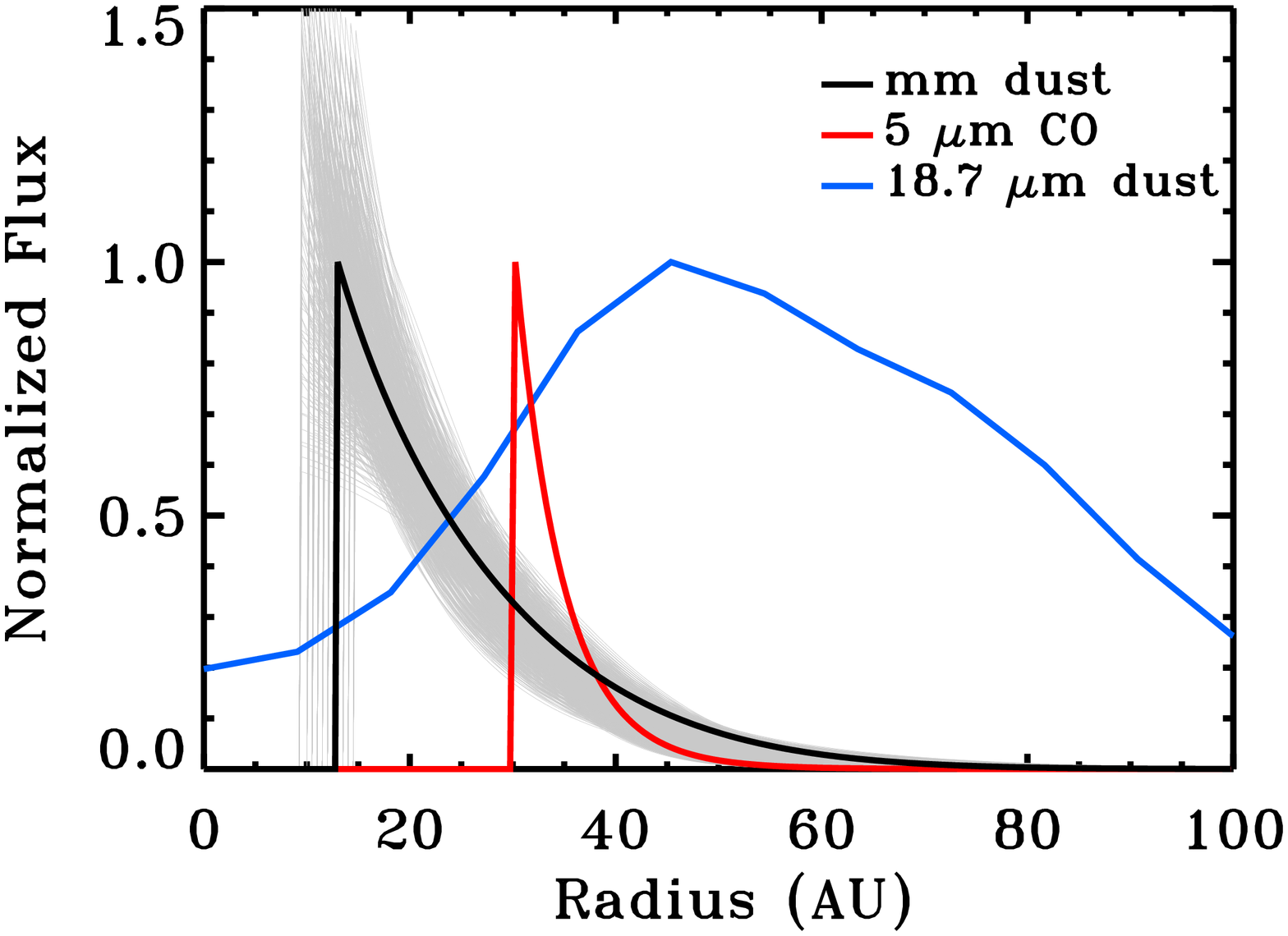} 
\end{center}
\caption{The radial distribution of flux from the millimeter dust
(black), 5 $\mu$m CO (red) and 18.7 $\mu$m dust (blue). The millimeter
profile is the best fit profile listed in Table \ref{table:model} with
the grey region representing fits within the error bars. The 5 $\mu$m
CO profile is the best fit model from \citet{brown12} Figure 8. The
18.7 $\mu$m dust flux is a cut along the major axis of the
\citet{geers07} VISIR image after 2-D maximum likelihood deconvolution
of the PSF. \label{fig:fluxprofiles}}
\end{figure}

\section{Discussion}
\label{discuss}

Oph IRS 48 is well-studied with resolved measurements of the disk and
central hole at several wavelengths and, unusually, constraints on
both the gas and dust. Despite spatially resolved rings of emission in
the 18.7 $\mu$m dust (55 AU peak; \citealt{geers07}) and the 5 $\mu$m
CO (30 AU peak; \citealt{brown12}), the inner edge of the millimeter
dust lies interior to both, with a cutoff in the dust density at the
even smaller radius of 12.9 $^{\rm +1.7}_{\rm -3.4}$ AU
(Fig. \ref{fig:fluxprofiles}). The SMA image definitively rules out
the millimeter emission following the structure in either the 5 $\mu$m
CO or the 18.7 $\mu$m dust. The central hole is barely resolved with
the SMA with a large portion of flux within the central beam. Thus,
all models require a significant percentage of the millimeter flux to
come from within the CO and IR dust rings. If the disk is currently
forming planets, the complex disk structure may provide
regions advantageous to planet and planetesimal formation. It may be
the case that already formed planets are responsible for some of the
sculpting seen in the disk.

Both infrared measurements likely trace only the surface layers of the
disk. The VLT-CRIRES 5 $\mu$m CO observations are very sensitive to
warm CO with constraints both spatially and spectrally. However, the
data are less sensitive to cold CO and completely insensitive to
non-molecular gas or other species of molecular gas. The VISIR 18.7
$\mu$m image has a PSF of $\sim$0\farcs5 which is similar to the CO
hole size. The 18.7 $\mu$m dust hole may be larger than the 30 AU
previously reported \citep{geers07} as seen in the deconvolved cut in
Figure \ref{fig:fluxprofiles} (noted also by Maaskant et al., in prep;
Bruderer et al., 2011, priv. comm.). While the majority of the 18.7
$\mu$m emission is from outside the millimeter dust, the inner regions
may not be entirely void of infrared emitting dust.

The millimeter dust emission places some constraints on the amount of
gas within the hole under assumptions of a canonical gas to dust
ratio. In the best fit model, 71\% of the millimeter flux arises from
dust within the 30 AU radius gas ring. From all the models, 40-90 \%
of the millimeter emission must be interior to 30 AU. We use the
\citet{beckwith90} conversion of millimeter flux to disk mass with a
$\beta$ of 1 \citep{rodmann06}. The dust temperature is taken to be in
the range of 50-150 K. The gas mass within the 30 AU CO ring is then
10$^{-4}$ - 10$^{-3}$ M$_\odot$ corresponding to a CO mass of
10$^{-7}$-10$^{-6}$ M$_\odot$, assuming a H$_2$/CO number ratio of
10$^4$. In a flared disk with a scale height set to the hydrostatic
pressure at 200 K, the density of CO is 10$^6$-10$^7$ cm$^{-3}$ or a
surface density of 10$^{19}$-10$^{20}$ cm$^{-2}$, comparable to a disk
with no cavity. Significantly warmer gas is likely to produce 5 $\mu$m
CO emission detectable in the CRIRES data.

There are three basic scenarios to reconcile the millimeter dust and
5 $\mu$m CO emission: 1.) the gas (and dust) is cold and therefore not
detectable in the infrared, 2.) the CO is photodissociated and only
atomic gas is left interior to 30 AU, or 3.) the gas has been removed
leaving millimeter dust and PAHs. 

\subsection{Scenario 1} 
Shadowing from a puffed up inner rim, either at the millimeter dust
wall or from atomic gas or PAHs closer to the star, could cool the
surface layers of the disk further out. The 5 $\mu$m CO limits the
amount of warm gas within the 30 AU gas radius but the limits are much
less stringent at cooler temperatures (\citealt{brown12}, their Figure
12). A temperature of $\sim$150 K or less throughout the inner disk is
needed to reconcile the 5 $\mu$m CO surface density limit with the
modeled millimeter surface density, assuming a gas-to-dust ratio of
100. Any illuminated wall would also likely need to be devoid of
CO. The shadowing region would require a large scale height to shadow
the disk out to 30 AU. In order for this scenario to be plausible,
the inner regions must be very settled while the outer structure is
flared, exposing the surface layers to stellar radiation to produce
the infrared emission.

\subsection{Scenario 2} 
Strong enough UV fields can photodissociate CO, with the
photodissociation rate in an unattenuated interstellar field being 2 x
10$^{-10}$ s$^{-1}$. The UV field around an A0 star such as IRS 48 is
much stronger than the interstellar field even in the critical
912-1100 \AA\, range and would lead to a photodissociation lifetime of
$<$1 year. However, CO photodissociation is subject to self-shielding
with the lines becoming saturated at a $^{12}$CO column density of
about 10$^{15}$ cm$^{-2}$, strongly decreasing the photodissociation
rate \citep{vandishoeck88, visser09}. Both millimeter dust grains and
PAH molecules are also effective absorbers of UV radiation, lowering
the UV field strength and thus the photodissociation rate. The
estimated CO density based on the millimeter dust would be strongly
self-shielding making complete photodissociation unlikely.

\subsection{Scenario 3} 
Removal of only the gas within 30 AU resulting in a strongly decreased
gas-to-dust ratio is a final possibility. Photoevaporation primarily
affects the gas with small dust grains tied to the gas flow. It may be
possible for large enough dust grains to remain. Destructive
collisions of the remaining large particles would then be needed to
generate the PAHs also seen from the inner regions.  \\ \\ 

These different scenarios should leave differing observational traces
which further observations may be able to distinguish. In the case of
cold CO, ALMA observations will have the sensitivity and spatial
resolution to look for cold CO within the central region of the
disk. For photodissociated CO, optical spectroastrometric or
high spectral resolution submillimeter observations of atomic C or O
would determine if the photodissociation products of CO are present
within 30 AU. If the gas is removed completely, neither cold CO nor
the atomic components of CO would be present in the inner disk region.

The ring of 18.7 $\mu$m dust is even more difficult to understand with
most of the emission coming from outside either the millimeter dust or
the 5 $\mu$m CO. The dominant 18.7 $\mu$m flux from $\sim$55 AU has no
obvious counterpart in the millimeter emission. Like the 5 $\mu$m CO,
the 18.7 $\mu$m flux comes from the disk surface and could reflect
differences in illumination and heating of the surface layers rather
than underlying changes in the distribution. However, the
centrally-peaked PAH emission points to some heating and small
particles in the inner region. Recent studies in the mid-IR and
scattered light suggest that small surface particles within the
cavities seen in larger grains are common \citep{muto12,
dong12}. Holes visible in the mid-IR are often of similar size to
those seen in the submillimeter (e.g. \citealt{panic12,
thalmann10}). However, the scattered light profiles, tracing the small
surface grains, are smooth with no decrease within the cavity. Dust
filtration or trapping creating grain size differences across
planetary gaps has been proposed \citep{rice06, pinilla12}, with the
largest particles left on the outside of the gap and smaller
particles, being more closely tied to the gas, allowed closer to the
star.  While the mid-IR dust and PAHs in Oph IRS 48 may fit this
paradigm, the large millimeter dust interior to the smaller mid-IR
emitting dust does not. Likewise, 5 $\mu$m CO emission usually arises
interior to submillimeter holes rather than from further out in disk
\citep{salyk09, pontoppidan08}. Continued multiwavelength studies are
vital to search for other examples of the inside-out nesting structure
seen in the Oph IRS 48 disk.

\section{Conclusions}

We have used new spatially resolved SMA 880 $\mu$m observations to
place stringent constraints of the dust distribution within the Oph
IRS 48 disk. The SMA data clearly rules out that the millimeter dust
distribution following either the 5 $\mu$m CO or 18.7 $\mu$m dust . A
large fraction of the millimeter flux is from dust interior to both
the 18.7 $\micron$ dust ring (55 AU radius) and the 5 $\mu$m CO
emission (30 AU radius). The millimeter dust also shows a strong drop
in surface density with a null in the visibilities. Our similarity
solution models place the hole at 12.9 $^{\rm +1.7}_{\rm -3.4}$
AU. The millimeter emission clearly shows that an empty cavity is not
responsible for the CO gas ring described in \citet{brown12}.  Further
observations are needed to distinguish between the potential scenarios
of cold shadowed gas, photodissociated CO and low gas-to-dust ratio,
but all these scenarios result in an unusual environment for potential
planet formation.




\acknowledgments {The Submillimeter Array is a joint project between
the Smithsonian Astrophysical Observatory and the Academia Sinica
Institute of Astronomy and Astrophysics and is funded by the
Smithsonian Institution and the Academia Sinica. The authors thank Jes
Jorgenesen for providing the earlier 230 GHz data and Vincent Geers for the
reduced VISIR data. J.M. Brown acknowledges the Smithsonian
Astrophysical Observatory for support from a SMA
fellowship. Astrochemistry at Leiden is supported by a Spinoza grant
from the Netherlands Organization for Scientific Research (NWO) and by
EU A-ERC grant 291141 CHEMPLAN. } Facilities: \facility{SMA}

\bibliographystyle{apj}

\end{document}